\documentclass[10pt]{article}
\usepackage[centertags]{amsmath}
\usepackage{amsfonts}
\usepackage{amssymb}
\usepackage{amsthm}
\usepackage{newlfont}
\usepackage[english]{babel}

\renewcommand\thesection{\arabic{section}.}

\begin{document}

\def\baselinestretch{1.7}
\title
{New String Theories And Their Generation Number}

\author{ Arel Genish and Doron Gepner\\
Department of Particle Physics\\
Weizmann Institute\\
Rehovot 76100, Israel}

\maketitle

\begin{abstract}
New heterotic string theories in four dimensions are constructed by tensoring a nonstandard SCFT along with some minimal SCFT's. All such theories are identified and their particle generation number is found. We prove that from the infinite number of new heterotic string theories only the \{6\} theory predicts three generations as seen in nature which makes it an interesting candidate for further study.

\end{abstract}

\smallskip

%%% ---------------------------------------------------------------------------
\goodbreak
\section{Introduction}

Heterotic string theories are build via the Gepner construction \cite{db3,db4} with the heterotic string map. The construction consists of several building blocks of specific central charges. For consistent theories in four dimension$(D=4)$ one of these building blocks is an $N=2$ SCFT with central charge nine \cite{db}.\\
In this paper, following reference \cite{db}, we realize the Gepner construction by choosing the $N=2$ SCFT as a tensor product of a nonstandard SCFT \cite{db} along with some minimal SCFT's \cite{Qiu}. The nonstandard SCFT model is specified by two strange integers, $m$ and $N$ where $m$ obeys, $N/2 < m < N$. We also tensor along with this theory $r$ minimal $N = 2$ SCFT's, specified by the integers $N_i$ where $i = 1, 2, . . . , r$.
The central charge of the complete theory is than given by,
\begin{eqnarray}\label{db4.1}
  9=c_{NS}+\sum^r_{i=1}c_i=\frac{3N}{2m-N}+\sum^r_{i=1}\frac{3N_i}{N_i+2}.
\end{eqnarray}
Solutions to this equations specify the $N=2$ SCFT content of consistent heterotic string theories in $D=4$. \\
The particle generation number of these string theories can be found from the spectrum of the $N=2$ SCFT as follows.
The string theories are described with a gauge group that includes $E_8\times E_6$ along with chiral fermions generations and anti generations in the $27$ and $\bar{27}$ representations of $E_6$ respectively. Particle generations correspond \cite{db4} to the $N=2$ SCFT $(C,C)$ fields of dimensions $h=\frac{1}{2}$ and
$U(1)$ charge $Q=1$. Where $(C,C)$ are left and right chiral super primary fields
satisfying $Q=\bar{Q}$. While anti generations come from $(C,A)$ fields with
$h=\frac{1}{2}$ and $Q=1$. Here, $(C,A)$ is a left chiral and right anti chiral super
primary such that $Q=-\bar{Q}$.
Thus we need to study the $(C,C/A)$ fields of the full theory.\\
We assume the diagonal modular invariant for all sub
theories thus there are no $(C,A)$ fields in any of the sub theories. For the $N=2$
minimal SCFT's we denote the $(C,C)$ fields as $\Phi^{N_i}_{k_i}$, their $U(1)$ charges were found in \cite{Qiu,Qiu7}. In \cite{db} the authors studied the left chiral super primary fields of the $(m,N)$ nonstandard $N=2$ SCFT, denoted $\rho^{m,N}_q(z,\bar{z})$.\\
We assumed the diagonal modular invariant for all sub theories, thus the only $(C,C)$ fields in the full theory are those constructed from $(C,C)$ fields of the sub theories
\begin{eqnarray}\label{db5.6}
 D_{\vec{k},q}=\prod^r_{i=1} \Phi^{N_i}_{k_i}\rho^{m,N}_q,
\end{eqnarray}
where $q=0,1\ldots,N$, and $k_i=0,1\ldots,N_i$. For our work we need $(C,C)$
fields whose total $U(1)$ charge is exactly $1$. These are the solutions of
the equation
\begin{eqnarray}\label{db5.9}
 1=Q=\sum^r_{i=1}Q^{N_i}_{k_i}+Q^{(m,N)}_q=\sum^r_{i=1}\frac{k_i}{N_i+2}+\frac{N}{2m-N}\left[\frac{mq}{N}\right].
\end{eqnarray}
Generations in the string theory are in one to one correspondence with
solutions of this equation, specified by $(k_1...k_r,q)$. For
the diagonal modular invariant there are no additional generations.\\
The anti generations of the theory come from $(C,A)$ fields with $Q=1$. For the diagonal modular invariant these fields are not present in the sub theories. However, the use of the Gepner construction breaks left right symmetry and such fields can appear in the full theory. By definition these fields have $Q=-\bar{Q}$ where $Q$ is given by eq. (\ref{db5.9}).
Thus clearly solutions of equation (\ref{db5.9}) correspond also to $(C,A)$ fields with
$Q=1$. However, the anti generations in the string theory are not in one to one
correspondence with these solutions. The Gepner construction provides a condition
resolving which of these fields correspond to an anti generation \cite{db}. Define the lattice $\hat{Q}$ as the lattice spanned by the $r+1$ vector $(-1,-1,\ldots,-1,1)$, where it is minus one for the first $r$ indices of the minimal models, and one for the last index of the non-standard model. Define also the lattice $K$ spanned by the vectors $\hat{N}_i\equiv N_i+2$ in the $i$'th position and zero everywhere else, for $i = 1,2,\ldots,r$, along with the vector $\hat{N}_{r+1}\equiv2m-N$ in the last position, and zero everywhere else. The anti generations are in one to one correspondence with solutions satisfying,
\begin{eqnarray}\label{db5.13}
\vec{k}\equiv(k_1,k_2...k_r,q+2r_q), \hspace{0.3in}  \vec{k}\in \hat{Q}+K,
\end{eqnarray}
where $r_q$ is an integer defined by
\begin{eqnarray}\label{db5.12}
  r_q \equiv \left[\frac{mq}{N}\right]-\frac{mq}{N}.
\end{eqnarray}
Here we defined $[x]$ as the fractional part of $x$, such that if $x$ is some integer
$[n\neq0]=1$ and $[0]=0$.
Finally, note that for non diagonal modular invariants there could be
additional anti generations.

%%% ---------------------------------------------------------------------------
\goodbreak
\section{String Theories in Dimension $D=4$}

One of our goals in this work is to find new string theories consistent in $D=4$. As said in the introduction these new string theories correspond to solutions of equation (\ref{db4.1}). These solutions are given by sets $\{N_r\}=\{N_1,\ldots,N_r\}$ specifying $r$ standard models and two strange integers $(m,N)$ specifying the nonstandard model. We will assume with no loss of generality that the set $\{N_r\}$ is arranged such that $N_{i+1}\geq N_i$. To identify $D=4$ consistent string theories we observe that $c_{NS}$ can be any rational number such that $3<c_{NS}\leq 9$ while $c_i$ are also rational numbers. Thus the condition
\begin{eqnarray}\label{a1}
  6>\sum^r_{i=1}\frac{3N_i}{N_i+2},
\end{eqnarray}
on the set $\{N_r\}$, is a necessary sufficient condition for a consistent $D=4$ string theory. Since $c_i$ are a rational numbers such that $1\leq c_i<3$ we have $\{N_r\}$ sets, $r \leq 4$, for which at least one of the $N_i$ is not restrained by condition (\ref{a1}). These correspond to an infinite number of new string theories consistent in $D=4$ which are given by,
\begin{eqnarray}\label{a2}
\{N_r\}=\begin{cases} \{N_1\} & N_1 \in Z^+   \\
\{N_1,N_2\} & N_1,N_2 \in Z^+  \\
\{1,N_2,N_3\} & N_2\leq 4,\; N_3 \in Z^+  \\
\{2,2,N_3\} & N_3 \in Z^+ \\
\{1,1,1,N_4\} & N_4 \in Z^+  \end{cases}.
\end{eqnarray}
while $(m,N)$ specifying the nonstandard model are given by equation (\ref{db4.1}). The infinite number of consistent $D=4$ string theories is a rich playing ground for phenomenological work on one hand. On the other hand, it makes the classification of all three generations models considerably more difficult.\\
In addition to the infinite number of theories given by eq. (\ref{a2}) we have $116$ consistent $D=4$ string theories, their sub theories content is given in the appendix. These consist of one $r=0$ theory, ninety nine theories with $r=3$, thirteen theories for which $r=4$ and three $r=5$ theories. Following our discussion, theories with $r>5$ are not consistent for $D=4$ thus we have found all the new $D=4$ consistent string theories containing one nonstandard model. Note that in our work we consider only theories containing one nonstandard model, other consistent $D=4$ theories can be constructed using two nonstandard models.\\

%%% ---------------------------------------------------------------------------
\goodbreak
\section{Generation Number}

We have already stated that we have found an infinite number of new string
theories consistent in $D=4$. Subsequently, to completely map all the new
string theories generation and anti generation numbers an analytical solution
must be found for the cases presented in eq. (\ref{a2}). We find analytical solutions for
the generation and anti generation numbers for all cases. We will present detailed solutions for the first two cases.
%\begin{eqnarray}\label{a12}
%\{N_r\}=\begin{cases} \{N_1\} & N_1 \in Z^+   \\
%\{N_1,N_2\} & N_1,N_2 \in Z^+  \end{cases}.
%\end{eqnarray}
Solutions for the other three cases are achieved in a similar fashion and we will
present them briefly. Using these solutions and a numerical solution for the $116$ cases appearing in the appendix we will find the $\{6\}(13,18)$ theory to be the only theory with three generations as seen in nature. Let us start with the $\{N_1\}$ theory, where $N_1 \in Z^+$.

\subsection{The $\{N_1\}$ Theory}

To solve the $\{N_1\}$ theory we study the central charge condition and the $U(1)$ charge
equation. First, note that when tensoring one standard model along with one nonstandard
model the central charge condition eq. (\ref{db4.1}) implies
\begin{eqnarray}\label{a13}
\nonumber \frac{m}{N}=\frac{3N_1+8}{4N_1+12}, \hspace{0.9in} \\
\Rightarrow gm=3N_1+8,   \hspace{0.5in} gN=4N_1+12.
\end{eqnarray}
Where since $m$ and $N$ are strange integers we have defined $g$, which is the
greatest common divisor (gcd) of $3N_1+8$ and $4N_1+12$.\\
We now prove that $g=1,2,4$ for $N_1=1+2n,2+4n,4n$ respectively. We can rewrite $m/N$ for
the different cases of $N_1$:
\begin{eqnarray}\label{a14}
\frac{m}{N}= \begin{cases} \frac{6n+11}{8n+16} & N_1=1+2n   \\
\frac{12n+14}{16n+20} & N_1=2+4n  \\
\frac{12n+8}{16n+12} & N_1=4n \end{cases}.
\end{eqnarray}
We immediately find that $g\geq1,2,4$ for the three cases respectively. To
prove that these are exact we examine each case. For example in the second
case assume that $g>2$, so that
\begin{eqnarray}\label{a15}
\nonumber \frac{m}{N}=\frac{12n+14}{16n+20}, \hspace{0.9in} \\
\Rightarrow \tilde{g}m=6n+7,   \hspace{0.5in} \tilde{g}N=8n+10,
\end{eqnarray}
where $g=2\tilde{g}$. Using these equations we can rewrite $\tilde{g}$ as
\begin{eqnarray}\label{a16}
     \tilde{g}= \frac{2}{3N-4m} \;\;\; \Rightarrow \;\;\; \tilde{g}\leq2,
\end{eqnarray}
clearly $\tilde{g}=2$ is not consistent since,
\begin{eqnarray}\label{a17}
     \tilde{g}=2 \;\;\; \Rightarrow \;\;\; m=3n+\frac{7}{2}\notin \texttt{Z}.
\end{eqnarray}
So we find $\tilde{g}=1$ thus $g=2$. Similar profs can be written for each case
and we find
\begin{eqnarray}\label{a18}
g= \begin{cases} 1 & N_1=1+2n   \\
2 & N_1=2+4n  \\
4 & N_1=4n   \end{cases}.
\end{eqnarray}
We now turn to the generation number, from eq. (\ref{a13}) we find the following relation
\begin{eqnarray}\label{a19}
2m-N=\frac{2}{g}(N_1+2).
\end{eqnarray}
Using this relation we can write the $U(1)$ charge equation (\ref{db5.9}) as
\begin{eqnarray}\label{a20}
1= \frac{2k_1+gN[\frac{qm}{N}]}{2(N_1+2)}.
\end{eqnarray}
Note that $q=0,1,\ldots,N$ while $m$ and $N$ are strange integers, this means that
$[\frac{qm}{N}]=0,\frac{1}{N},\frac{2}{N},\ldots,1$ albeit not respectively. We can now
count solutions for the different cases of $N_1$. First, note that for all cases
$N=\frac{4}{g}(N_1+3)$ while $N[\frac{qm}{N}]=0,1,2,\ldots,N$. Thus for the $g=1,2$ cases
for every $k_1=0,1,\ldots,N_1$ we can choose some $q=0,1,\ldots,N$ such that the $U(1)$
charge equation is satisfied. The number of solutions is the generation number denoted
$G$ and is thus given for all $g=1,2$ models by
\begin{eqnarray}\label{a21}
    G=N_1+1.
\end{eqnarray}
For the $g=4$ case the $U(1)$ charge equation becomes
\begin{eqnarray}\label{a22}
1= \frac{k_1+2N[\frac{qm}{N}]}{N_1+2}.
\end{eqnarray}
Here, $N_1=4n$ so that $N_1+2$ is an even number while $2N[\frac{qm}{N}]$ is an even
number, as well. Thus for the $g=4$ case solutions are found only for even $k_1$ and the
generation number is given by
\begin{eqnarray}\label{a23}
    G=\frac{N_1}{2}+1.
\end{eqnarray}
To conclude, the generation number for the $\{N_1\}$ theory is given by
\begin{eqnarray}\label{a24}
G= \begin{cases} N_1+1 & N_1=1+2n   \\
N_1+1 & N_1=2+4n  \\
\frac{N_1}{2}+1 & N_1=4n   \end{cases}.
\end{eqnarray}
To find the anti generation number, denoted $\bar{G}$, rewrite the $U(1)$
charge equation (\ref{a22}) using eq. (\ref{db5.12}) as
\begin{eqnarray}\label{a25}
1= \frac{2k_1+gmq+gNr_q}{2(N_1+2)}.
\end{eqnarray}
Next, with the aid of eq. (\ref{a13}) and the definition of $k_2=q+2r_q$ this
equation can be brought to the form
\begin{eqnarray}\label{a26}
k_1+k_2=(N_1+2)(1-\frac{3q}{2}-2r_q).
\end{eqnarray}
We now briefly turn to the lattice $\hat{Q}+K$. The lattice $\hat{Q}+K$ is
spanned by the three vectors
\begin{eqnarray}\label{a27}
 v_0=(-1,1), \;\;  v_1=(N_1+2,0), \;\; v_2=(0,2m-N).
\end{eqnarray}
Since $g$ is either one or an even integer, we observe from equation
(\ref{a19}) that $|v_l|=a|v_s|$. Here $a$ is a positive integer and $v_s$
($v_l$) stands for the shorter (longer) of the two vectors $v_1$ and $v_2$.
Thus the vector $\vec{k}$ lies in the lattice if and only if
\begin{eqnarray}\label{a28}
     k_1+k_2=0 \;\;\;  \text{mod } |v_s|.
\end{eqnarray}
Finally, using this condition and equation (\ref{a26}) we can identify the
theory anti generations. For the $N_1=2n+1$ case observe that the left hand
side of eq. (\ref{a26}) is an integer. Since $N_1$ is odd, the right hand side of
eq. (\ref{a26}) is an integer only for $q$ even so all the solutions of the $U(1)$
charge equation are such that $q$ is even. Next, since $q$ is even and
$|v_s|=N_1+2$ we find from eq. (\ref{a26}) that all the solutions satisfy the
condition (\ref{a28}). We have thus shown that all the solutions of the $U(1)$
charge equation lay in the $\hat{Q}+K$ lattice so that the anti generations
number $\bar{G}$ is given by,
\begin{eqnarray}\label{a29}
    \bar{G}=N_1+1.
\end{eqnarray}
For the $g=2$ case $|v_s|=N_1+2$. Here $N_1=2+4n$ is even so that the RHS of
equation (\ref{a26}) is an integer irrespective of the $q$ value. Solutions
with even $q$ satisfy the condition (\ref{a28}) and thus lie in the $\hat{Q}+K$
lattice. To identify solutions with even $q$ we can write eq. (\ref{a26}) using
$N_1=2+4n$
\begin{eqnarray}\label{a30}
k_1+k_2=2(n+1)(2-3q-4r_q).
\end{eqnarray}
Clearly all the solutions satisfy $k_1+k_2$ even so that $k_1$ and $k_2$ are both even or
both odd. Since $k_2=q+2r_q$ the parity of $k_2$ is the same as the parity of $q$ we thus
conclude that solutions of even $k_1$ satisfy the condition (\ref{a28}). The anti
generation number is then given by
\begin{eqnarray}\label{a31}
    \bar{G}=\frac{N_1}{2}+1.
\end{eqnarray}
For the $g=4$ case $N_1=4n$ so we find $|v_s|=2m-N=2n+1$ and eq. (\ref{a26}) can be
written as
\begin{eqnarray}\label{a32}
k_1+k_2=2(2n+1)(1-\frac{3q}{2}-2r_q).
\end{eqnarray}
Thus we find that all solutions satisfy the condition (\ref{a28}). Finally, the
anti generation number is given by
\begin{eqnarray}\label{a33}
\bar{G}= \begin{cases} N_1+1 & N_1=1+2n   \\
\frac{N_1}{2}+1 & N_1=2+4n  \\
\frac{N_1}{2}+1 & N_1=4n   \end{cases}.
\end{eqnarray}
The net number of generations $G_{net}$ is the difference between the
generation and anti generation numbers and can now be found for the different
cases
\begin{eqnarray}\label{a34}
G_{net}=G-\bar{G}= \begin{cases} 0 & N_1=1+2n   \\
\frac{N_1}{2} & N_1=2+4n  \\
0 & N_1=4n   \end{cases}.
\end{eqnarray}
To conclude, note that the only model corresponding to three net generations is the
$\{6\}(13,18)$ theory.

\subsection{The $\{N_1,N_2\}$ Theory}

To find the $\{N_1,N_2\}$ theory generation and anti generation numbers $G$ and $\bar{G}$
we first study the central charge equation,
\begin{eqnarray}\label{d1}
3=\frac{\hat{N}_1-2}{\hat{N}_1}+\frac{\hat{N}_2-2}{\hat{N}_2}+\frac{N}{2m-N}.
\end{eqnarray}
Where we have used the notation $\hat{N}_i=N_i+2$ introduced above. From this
equation we find $m$ and $N$ to be
\begin{eqnarray}\label{d2}
\nonumber \frac{m}{N}=\frac{\hat{N}_1\hat{N}_2+\hat{N}_1+\hat{N}_2}{\hat{N}_1\hat{N}_2+2\hat{N}_1+2\hat{N}_2},
\end{eqnarray}
\begin{eqnarray}
\Rightarrow gm=\hat{N}_1\hat{N}_2+\hat{N}_1+\hat{N}_2,   \hspace{0.3in}  gN=\hat{N}_1\hat{N}_2+2\hat{N}_1+2\hat{N}_2.
\end{eqnarray}
In the proceeding we will frequently use the following relations implied by
the solutions for $m$ and $N$,
\begin{eqnarray}\label{d3}
g(N-m)=\hat{N}_1+\hat{N}_2,   \hspace{0.3in}   g(2m-N)=g\hat{N}_3=\hat{N}_1\hat{N}_2.
\end{eqnarray}
In addition all the variables introduced are positive
integers.\\
We now examine the implications of the above solutions on $\hat{N}_1$ and
$\hat{N}_2$, we first define
\begin{eqnarray}\label{d4}
 \hat{N}_1=hc,  \hspace{0.2in} g=hd,   \hspace{0.4in} \hat{N}_2=le,   \hspace{0.2in} g=lf.
\end{eqnarray}
Here, $h$($l$) is the $gcd$ of $g$ and $\hat{N}_1$($\hat{N}_2$) which fixes
$c$ and $d$($e$ and $f$) as strange integers. Note that our equations are
$\hat{N_1} \leftrightarrow \hat{N_2}$ symmetric this implies that $h=l$ and
$f=d$. Clearly it is sufficient to prove that $h=l$, we write $\hat{N}_1/g$
using eq. (\ref{d3})
\begin{eqnarray}\label{d5}
 \frac{\hat{N}_1}{g}=\frac{g(N-m)-\hat{N}_2}{g}=\frac{l\left(f(N-m)-e\right)}{lf}.
\end{eqnarray}
Thus $l$ is a common divisor of $\hat{N}_1$ and $g$. Since $h$ is the gcd of
$\hat{N}_1$ and $g$ we find $h\geq l$. Due to the $\hat{N}_1 \leftrightarrow
\hat{N}_2$ symmetry the same argument can be made for $\hat{N}_2/g$ from
which follows $l\geq h$. Finally, to satisfy both conditions $h=l$.\\
Another important implication of eq. (\ref{d3}) follows from the observation that
$\frac{\hat{N}^2_1}{g}=hc\frac{c}{d}$ is an integer,
\begin{eqnarray}\label{d6}
 \hat{N}^2_1=g\hat{N}_1(N-m)-\hat{N}_1\hat{N}_2=g\left(\hat{N}_1(N-m)-\hat{N}_3\right).
\end{eqnarray}
Considering that $c$ and $d$ are strange integers, $hc\frac{c}{d}$
can only be an integer if $hc=ad$. Again, since $c$ and $d$ are strange
integers the last equality sets $h=nd$. To conclude, we have found that
\begin{eqnarray}\label{d7}
\nonumber \hat{N}_1=ndc,\hspace{0.2in} \hat{N}_2=nde,\hspace{0.2in} \hat{N}_3=nce,\hspace{0.2in} g=nd^2,   \\
  m=nce+\frac{c+e}{d},   \hspace{0.3in}   N=nce+2\frac{c+e}{d}.\hspace{0.3in}
\end{eqnarray}
Finally, we prove that $c$ and $e$ are strange integers. Assume that $c$ and
$e$ are not strange. We can always define $c=\gamma a$ and $e=\gamma b$ such
that $\gamma$ is the gcd of $c$ and $e$ thus $a$ and $b$ are strange
integers. We have defined $c$ and $d$ as strange integers it follows that
$\gamma$ and $d$ are strange as well. Next, from eq. (\ref{d7}) we observe that
$\frac{\gamma(a+b)}{d}$ is an integer so we can write $pd=\gamma(a+b)$.
Since $\gamma$ and $d$ are strange $p=l\gamma$ and $c+e=l\gamma d$. Using
these definitions and eq. (\ref{d7}) we can write $m$ and $N$ as
\begin{eqnarray}\label{d8}
m=nab\gamma^2+\gamma l,  \hspace{0.4in}  N=nab\gamma^2+2\gamma l.
\end{eqnarray}
Thus $\gamma$ is a common divisor of $m$ and $N$. On the other hand, since $m$
and $N$ are strange, their only common divisor is one so $\gamma=1$,
consequently, $c$ and $e$ are strange.\\
We now turn to the $U(1)$ charge equation (\ref{db5.9}). Using eqs. (\ref{d3},\ref{d7}) it can be written as
\begin{eqnarray}\label{d9}
1=\frac{k_1\hat{N}_2+k_2\hat{N}_1+gN[\frac{qm}{N}]}{\hat{N}_1\hat{N}_2}=
\frac{k_1e+k_2c+dN[\frac{qm}{N}]}{ndce}.
\end{eqnarray}
Note that the denominator and the last term in the numerator are integer
multiples of $d$. To satisfy the $U(1)$ charge equation the first two terms
in the numerator should also be an integer multiple of $d$. This condition
can be written in the following manner
\begin{eqnarray}\label{d10}
k_1(e+c)+c(k_2-k_1)=ad.
\end{eqnarray}
If we now divide by $d$ the RHS and the first term on the LHS are clearly
integers(see eq. \ref{d7}). From the remaining term we see that solutions of the $U(1)$ charge
equation obey
\begin{eqnarray}\label{d11}
k_1-k_2=ld \;\;\;\;  \Rightarrow  \;\;\;\;
\left\lceil-\frac{k_2}{d}\right\rceil \leq l\leq \left\lfloor nc-\frac{2+k_2}{d}\right\rfloor.
\end{eqnarray}
Here, we have used $k_1=0,1,\ldots,\hat{N}_1-2$ and since $l$ is an integer we
introduced the notation $\lfloor x\rfloor/ \lceil x\rceil\equiv$ round
down/up $x$. Using this the $U(1)$ charge equation (\ref{d9}) can be written as
\begin{eqnarray}\label{d12}
1=\frac{k_2(c+e)+led+dN[\frac{qm}{N}]}{ndce}.
\end{eqnarray}
To count the solutions of this equation first note that
$N[\frac{qm}{N}]=0,1,\ldots,N$ and $N=nce+2\frac{c+e}{d}$. Thus for every
$k_2$ and $l$ which obey
\begin{eqnarray}\label{d13}
k_2(c+e)+led \leq ndce,
\end{eqnarray}
exits a solution. Combining this with eq. (\ref{d11}) we get the following limits for
$l$
\begin{eqnarray}\label{d14}
 l_{min}=\left\lceil-\frac{k_2}{d}\right\rceil, \hspace{0.3in}
l_{max}=\lfloor
Min\left(nc-\frac{k_2}{d}(1+\frac{c}{e}),nc-\frac{k_2+2}{d}\right)\rfloor.
\end{eqnarray}
The number of solutions to the $U(1)$ equation $G$ is thus given by
\begin{eqnarray}\label{d15}
G=\sum^{\hat{N}_2-2}_{k_2=0}(l_{max}-l_{min}+1).
\end{eqnarray}
To resolve the dilemma of $l_{max}$ we define $s=\lceil\frac{e}{c}\rceil$ for
$c\neq1$ so that
\begin{eqnarray}\label{d16}
l_{max}= \begin{cases}
\lfloor nc-\frac{k_2+2}{d}\rfloor & k_2\leq s-1  \\
\lfloor nc-\frac{k_2}{d}(1+\frac{c}{e})\rfloor & k_2\geq s   \end{cases}.
\end{eqnarray}
For $c=1$ we should shift $s$ by one. The error caused by using eq. (\ref{d16}) for
$c=1$ is
\begin{eqnarray}\label{d17}
\left\lfloor nc-\frac{s+2}{d}\right\rfloor-\left\lfloor nc-\frac{s}{d}(1+\frac{c}{e})\right\rfloor
=-\left\lceil\frac{e+2}{d}\right\rceil+\left\lceil\frac{e+1}{d}\right\rceil=-\delta_{e-d\lfloor\frac{e}{d}\rfloor,d-1}.
\end{eqnarray}
We can now write the generation number $G$ as the following sum
\begin{eqnarray}\label{d18}
\nonumber G=\hat{N}_2-1+\sum^{s-1}_{k_2=0}\left\lfloor nc-\frac{k_2+2}{d}\right\rfloor
+\sum^{\hat{N}_2-2}_{k_2=s}\left\lfloor nc-\frac{k_2}{d}(1+\frac{c}{e})\right\rfloor
-\sum^{\hat{N}_2-2}_{k_2=0}\left\lceil-\frac{k_2}{d}\right\rceil-c_1.\\
\end{eqnarray}
Where $c_1=\delta_{c,1}\delta_{e-d\lfloor\frac{e}{d}\rfloor,d-1}$ and it
compensates for the error in $l_{max}$. Before we solve this sum note that it
can be simplified by solving the sums over $nc$ and using
$\lfloor-x\rfloor=-\lceil x\rceil$,
\begin{eqnarray}\label{d19}
\nonumber G=\hat{N}_2-1+nc(\hat{N}_2-1)
+\sum^{\hat{N}_2-2}_{k_2=0}\left\lfloor\frac{k_2}{d}\right\rfloor
-\sum^{\hat{N}_2-2}_{k_2=s}\left\lceil\frac{k_2}{d}(1+\frac{c}{e})\right\rceil
-\sum^{s-1}_{k_2=0}\left\lceil\frac{k_2+2}{d}\right\rceil
-c_1.\\
\end{eqnarray}
The first sum can be solved as follows
\begin{eqnarray}\label{d20}
\sum^{\hat{N}_2-2}_{k_2=0}\left\lfloor\frac{k_2}{d}\right\rfloor=-ne+\left\lceil\frac{1}{d}\right\rceil+d\sum^{ne-1}_{k_2=0}k_2
=-ne+\left\lceil\frac{1}{d}\right\rceil+\frac{\hat{N}_2}{2}(ne-1).
\end{eqnarray}
To solve the second sum in eq. (\ref{d19}) we first solve the following sum
\begin{eqnarray}\label{d21}
\sum^{\hat{N}_2-2}_{k_2=1}\left\lceil\frac{k_2}{d}(1+\frac{c}{e})\right\rceil
=-2n(e+c)+\left\lfloor\frac{e+c}{ed}\right\rfloor+\sum^{\hat{N}_2}_{k_2=1}\left\lceil\frac{k_2(e+c)}{de}\right\rceil
\end{eqnarray}
If we look back to eq. (\ref{d7}) we see that since $m$ and $N$ are strange $\frac{e+c}{d}$ and
$e$ are strange as well. To solve the last sum in eq. (\ref{d21}) we first recall that if $a$
and $b$ are strange then $[\frac{k_2a}{b}]=\frac{1}{b},\frac{2}{b},\ldots,1$ for
$k_2=1,2,\ldots,b$. Next, for $k_2\neq 0$ we can write
$\lceil\frac{k_2(e+c)}{de}\rceil=\frac{k_2(e+c)}{de}+1-[\frac{k_2(e+c)}{de}]$. If we use
this to solve the last sum in eq. (\ref{d21}) we get
\begin{eqnarray}\label{d22}
\sum^{\hat{N}_2}_{k_2=1}\left(\frac{k_2(e+c)}{de}+1\right)-nd\sum^{e}_{k_2=1}\frac{k_2}{e}
=\frac{\hat{N}_2(ne+nc)+en+nc+\hat{N}_2-nd}{2}.
\end{eqnarray}
To solve the sum in eq. (\ref{d19}) we also need to solve
\begin{eqnarray}\label{d23}
\sum^{s-1}_{k_2=1}\left\lceil\frac{k_2(1+\frac{c}{e})}{d}\right\rceil=\sum^{s-1}_{k_2=1}\left\lceil\frac{k_2}{d}\right\rceil+a
\end{eqnarray}
Here we have defined $s-1=da+b$, where $a=\lfloor\frac{s-1}{d}\rfloor$ and
$b\leq d-1$. This equality can be explained as follows. We observe that $s$
was defined as the smallest integer for which $\frac{sc}{e}\geq1$. This means
that for $k_2\leq s-1$ the second term in the numerator $\frac{k_2c}{e}$ is
smaller than one. The first term $k_2$ is clearly an integer thus
\begin{eqnarray}\label{d24}
\left\lceil\frac{k_2(1+\frac{c}{e})}{d}\right\rceil= \begin{cases} \lceil\frac{k_2}{d}\rceil & k_2\neq nd \\
\lceil\frac{k_2}{d}\rceil+1 & k_2=nd \end{cases}.
\end{eqnarray}
With $n$ some positive integer. Finally, under the summation $k_2=nd$
exactly $a$ times. The sum on the RHS of eq. (\ref{d23}) can be solved, however if
we shift the last sum in eq. (\ref{d19}), we get
\begin{eqnarray}\label{d25}
\nonumber \sum^{s-1}_{k_2=0}\left\lceil\frac{2+k_2}{d}\right\rceil=\sum^{s-1}_{k_2=1}\left\lceil\frac{k_2}{d}\right\rceil+\left\lceil\frac{s+1}{d}\right\rceil+\left\lceil\frac{s}{d}\right\rceil-\left\lceil\frac{1}{d}\right\rceil
=\sum^{s-1}_{k_2=1}\left\lceil\frac{k_2}{d}\right\rceil+2a+1+\delta_{b,d-1}.\\
\end{eqnarray}
Here, in the last equality we have used the definition $s-1=ad+b$ and $b\leq d-1$.
Carefully gathering all the sums we get the generation number for the $\{N_1,N_2\}$
theory
\begin{eqnarray}\label{d26}
\nonumber G=\frac{n^2ecd+en+nc+nd}{2}-1-a-\left\lfloor\frac{e+c}{ed}\right\rfloor-\delta_{b,d-1}-c_1\\
=\frac{nd}{2}(m+1)-1-\left\lfloor\frac{s-1}{d}\right\rfloor-\left\lfloor\frac{e+c}{ed}\right\rfloor-\delta_{b,d-1}-c_1
\end{eqnarray}
Where we have used eq. (\ref{d7}) to simplify the result.\\
To find the anti generation number we first examine the $\hat{Q}+K$ lattice.
The $\hat{Q}+K$ lattice is spanned by
\begin{eqnarray}\label{d27}
 v_0=(-1,-1,1), \;\;  v_1=(cnd,0,0), \;\; v_2=(0,end,0), \;\; v_3=(0,0,nce).
\end{eqnarray}
A solution vector $\vec{k}$ lies in the lattice if it can be written as
\begin{eqnarray}\label{d28}
 (k_1,k_2,k_3)=\alpha(-1,-1,1)+\beta(cnd,0,0)+\phi(0,end,0)+\gamma(0,0,nce),
\end{eqnarray}
with integer coefficients. This leads to the following restrictions
\begin{eqnarray}\label{d29}
 \frac{k_1-k_2}{nd}=\beta c-\phi e, \hspace{0.15in}  \frac{k_2+k_3}{ne}=\gamma c+\phi d,
\hspace{0.15in}  \frac{k_1+k_3}{nc}=\gamma e+\beta d.
\end{eqnarray}
We observe that if a solution lies on the lattice it satisfies
\begin{eqnarray}\label{d30}
k_1-k_2=0 \text{ mod } nd.
\end{eqnarray}
We now show that if a solution satisfies this condition then it lies on the lattice, i.e
it satisfies eq. (\ref{d29}). First, we write the $U(1)$ equation (\ref{d9}) using eqs. (\ref{db5.12},\ref{d7}):
\begin{eqnarray}\label{d31}
\frac{k_1e+k_2c+k_3(c+e)}{ndce}=\zeta.
\end{eqnarray}
Where $\zeta=1-q-r_q$. This equation can be written as,
\begin{eqnarray}\label{d32}
\frac{k_1-k_2}{nd}+\frac{(k_2+k_3)(e+c)}{nde}=\zeta c,  \hspace{0.1in} \text{or} \hspace{0.2in} \frac{k_1+k_3}{nc}+\frac{k_2+k_3}{ne}=\zeta d.
\end{eqnarray}
If we now assume the condition (\ref{d30}), then the second term in the first equation is
an integer. From eq. (\ref{d7}) we note that $\frac{e+c}{d}$ and $ne$ are strange, thus we find
that
\begin{eqnarray}\label{d33}
k_2+k_3=0 \text{ mod } ne.
\end{eqnarray}
Using this the second equation leads to
\begin{eqnarray}\label{d34}
k_1+k_3=0 \text{ mod } nc.
\end{eqnarray}
To show that the point satisfies eq. (\ref{d29}), we note that since $d$ and $c$
are strange any integer can be written as $l=\phi d+\gamma c$. Thus if
eq. (\ref{d33}) is satisfied we can choose $\phi$ and $\gamma$ so that
\begin{eqnarray}\label{d35}
\frac{k_2+k_3}{ne}=\phi d+\gamma c.
\end{eqnarray}
We first use this in the second equation in (\ref{d32}), finding that
\begin{eqnarray}\label{d35.1}
\frac{k_1+k_3}{nc}=d(\zeta-\phi)-c\gamma.
\end{eqnarray}
Next, with the definition $\frac{c+e}{d}=p$ \, eq. (\ref{d35.1}) can be written as
\begin{eqnarray}\label{d35.2}
\frac{k_1+k_3}{nc}=(\zeta-p\gamma-\phi)d+\gamma e.
\end{eqnarray}
Finally, we use eq. (\ref{d35}) in the first equation in (\ref{d32}) to find
\begin{eqnarray}\label{d35.3}
\frac{k_1-k_2}{nd}=\zeta c-\phi dp-\gamma cp=(\zeta-p\gamma-\phi)c-\phi e.
\end{eqnarray}
So that if we choose $\beta=\zeta-p\gamma-\phi$ we find an integer solution
for all coefficients. Thus we have shown that a solution lies in the lattice
if and only if it satisfies eq. (\ref{d30}).\\
We can now count the solutions which satisfy eq. (\ref{d30}) in a similar manner as
previously, we get
\begin{eqnarray}\label{d36}
 \bar{G}=\hat{N}_2-1+\sum^{s-1}_{k_2=0}\left\lfloor c-\frac{k_2+2}{nd}\right\rfloor
+\sum^{\hat{N}_2-2}_{k_2=s}\left\lfloor c-\frac{k_2}{nd}(1+\frac{c}{e})\right\rfloor
-\sum^{\hat{N}_2-2}_{k_2=0}\left\lceil-\frac{k_2}{nd}\right\rceil-\bar{c}_1.
\end{eqnarray}
Where $\bar{c}_1$ is defined as
$\delta_{cn,1}\delta_{e-nd\lfloor\frac{e}{nd}\rfloor,nd-1}$. The sums in eq. (\ref{d36}) are
solved in a similar fashion to the sums appearing in the generation number. We find the
anti generation number for the $\{N_1,N_2\}$ theory,
\begin{eqnarray}\label{d37}
\bar{G}=\frac{d}{2}(m+1)-1-\bar{a}-\left\lfloor\frac{e+c}{edn}\right\rfloor-\delta_{b,nd-1}-\bar{c}_1,
\end{eqnarray}
where $\bar{a}$ is given by $\lfloor\frac{s-1}{nd}\rfloor$.\\
Finally, the net number of generation is given by $G-\bar{G}$. We now prove that the
$\{N_1,N_2\}$ theory does not contain a model with three net generations. First, using
$1-\frac{1}{n}\geq\frac{1}{n}\lfloor x\rfloor-\lfloor \frac{x}{n}\rfloor\geq0$ we find
the following upper bound for the anti generation number,
\begin{eqnarray}\label{d38}
\frac{G+1}{n}+1>\bar{G}.
\end{eqnarray}
This relation implies that all models with $G\geq10$ and $n\geq2$ will have a net
generation number bigger than three. Next, we observe that $G_{net}=0$ for all models with
$n=1$. Finally, by noting that $\left\lfloor\frac{e+c}{ed}\right\rfloor\leq1+\delta_{ecd,1}$ and
$a\leq\frac{s-1}{d}$ we get the following lower bound for $G$,
\begin{eqnarray}\label{d38}
\nonumber G\geq \frac{n^2ecd+nc+nd+en}{2}-\frac{1}{d}\left\lceil \frac{e}{c}\right\rceil+\frac{1}{d}-4-\delta_{ecd,1},
\end{eqnarray}
which is a monotonically increasing function of $c,e,d$ and $n$. Using this
lower bound we find that the only theories with $n\geq2$ which have less than
ten generations are the $\{1,1\}$ and $\{2,2\}$ theories with
$(n,c,e,d)=(3,1,1,1),(2,1,1,2)$, respectively. This can be verified by trying
to minimally increase $c,e,d$ or $n$ of the mentioned theories in accordance
with the restrictions found above and checking the lower bound for $G$. To
conclude, note that the net generation number for the $\{1,1\}$ and $\{2,2\}$
theories is two, so that no $\{N_1,N_2\}$ theory has three net generations.

\subsection{The $\{2,2,N_3\}$ Theory}

The $\{2,2,N_3\}$ theory is solved in a similar manner to the $\{N_1\}$ theory, from the
central charge equation we find
\begin{eqnarray}\label{b1}
\nonumber \frac{m}{N}=\frac{N_3+3}{N_3+4}, \hspace{0.9in} \\
\Rightarrow m=N_3+3,   \hspace{0.5in} N=N_3+4.
\end{eqnarray}
Here, the result for $\frac{m}{N}$ implies that $g=1$. Using the results for
$m$ and $N$ it is evident that
\begin{eqnarray}\label{b2}
\hat{N}_4=2m-N=N_3+2=\hat{N}_3.
\end{eqnarray}
Where we have used the notation $\hat{N}_i=N_i+2$. The $U(1)$ charge equation
can be written as
\begin{eqnarray}\label{b3}
1-\frac{k_1+k_2}{4}= \frac{k_3+N[\frac{qm}{N}]}{\hat{N}_3}.
\end{eqnarray}
It follows that all solutions obey $k_1+k_2=0$ mod $M_2$, where
$\frac{M_2}{M_3}=\frac{\hat{N}_2}{\hat{N}_3}$ so that $M_2$ and $M_3$ are strange. The
number of solutions corresponding to $(k_1,k_2)$ denoted $G_{k_1,k_2}$ can be found in a
similar way to the $\{N_1\}$ theory, equation (\ref{b3}) has a solution\footnote{An
exception arises when $\hat{N}_3=4$ and $k_1+k_2=1$ for which
$Min(\hat{N}_3(1-\frac{k_1+k_2}{4})+1,\hat{N}_3-2)=\hat{N}_3-2$.}
 for any $k_3$ up to
\begin{eqnarray}\label{b4}
G_{k_1,k_2}=\delta_{k_1+k_2,nM_2}\left( \hat{N}_3(1-\frac{k_1+k_2}{4})\pm1\right).
\end{eqnarray}
Where it is minus one for $(k_1,k_2)=(0,0)$ and plus one for all the rest
while the delta takes care of the restriction $k_1+k_2=0$ mod $M_2$. The
generation number can be written as
\begin{eqnarray}\label{b5}
G=\sum^{N_1}_{k_1=0}\sum^{N_2}_{k_2=0}G_{k_1,k_2}.
\end{eqnarray}
For the case in question $\hat{N}_2=4$, so that clearly $M_2=4,2,1$ for
$\hat{N}_3=1+2n,2+2n,4n$ respectively. Solving eq. (\ref{b5}) we find the generation number for
the $\{2,2,N_3\}$ theory
\begin{eqnarray}\label{b6}
G= \begin{cases} \hat{N}_3 & \hat{N}_3=1+2n   \\
\frac{5}{2}\hat{N}_3+3 & \hat{N}_3=2+4n  \\
\frac{9}{2}\hat{N}_3+7-2\delta_{\hat{N}_3,4} & \hat{N}_3=4n   \end{cases}.
\end{eqnarray}
To find the anti generation number we study the restrictions set by the demand $\vec{k}
\in \hat{Q}+K$ and the $U(1)$ charge equation (\ref{b3}). In a similar manner to the
$\{N_1,N_2\}$ theory we find that a solution lies in the $\hat{Q}+K$ lattice if and only
if
\begin{eqnarray}\label{b7}
 k_1+k_2=0 \text{ mod } 4, \hspace{0.15in}  k_3-k_2=0 \text{ mod } h .
\end{eqnarray}
Where $h=1,2,4$ is the gcd of $\hat{N}_2=4$ and $\hat{N}_3=1+2n,2+2n,4n$
respectively. The first restriction means that the only solutions that may
contribute to the anti generation number are $G_{0,0}$ and $G_{2,2}$. Using
the second restriction we find the anti generation number is given by
\begin{eqnarray}\label{b8}
\bar{G}=\sum^{2}_{i=1}\left(\left\lfloor\frac{G_{i}-1-a_i}{h}\right\rfloor+1\right),
\end{eqnarray}
where $G_i=G_{0,0},G_{2,2}$ and $a_i=k_2$ mod $h$, is the lowest $k_3$ satisfying the
restriction (\ref{b7}). Solving these sums for the different cases we get the anti
generation number for the $\{2,2,N_3\}$ theory
\begin{eqnarray}\label{b9}
\bar{G}= \begin{cases} \hat{N}_3 & \hat{N}_3=1+2n   \\
\frac{\hat{N}_3}{2}+1 & \hat{N}_3=2+4n  \\
\frac{\hat{N}_3}{4} & \hat{N}_3=4n   \end{cases}.
\end{eqnarray}
The net number of generation is easily calculated and clearly no $\{2,2,N_3\}$ theory
will produce three net generations.

\subsection{The $\{1,1,1,N_4\}$ Theory}

This theory is solved in the same way as the previous case. From the $U(1)$
charge equation
\begin{eqnarray}\label{c1}
1-\frac{k_1+k_2+k_3}{3}= \frac{k_4+N[\frac{qm}{N}]}{\hat{N}_4},
\end{eqnarray}
we get the restriction $k_1+k_2+k_3=0$ mod $M_2$. Where
$\frac{M_2}{M_4}=\frac{\hat{N}_1}{\hat{N}_4}$ so that $M_2$ and $M_4$ are strange. The
generation number is then given by
\begin{eqnarray}\label{c2}
G=\sum^{N_1}_{k_1=0}\sum^{N_2}_{k_2=0}\sum^{N_3}_{k_3=0}G_{k_1,k_2,k_3}.
\end{eqnarray}
Where
\begin{eqnarray}\label{c3}
G_{k_1,k_2,k_3}=\delta_{k_1+k_2+k_3,nM_2}\left( \hat{N}_4(1-\frac{k_1+k_2+k_3}{3})\pm1\right).
\end{eqnarray}
Here, it is minus one for $(k_1,k_2,k_3)=(0,0,0)$ and plus one for all the
rest\footnote{An exception arises for $\hat{N}_4=3$ and $k_1+k_2+k_3=1$.}.
Since $\hat{N}_2=3$ we get $\hat{M}_2=1,3$ for $\hat{N}_4=3n$ and
$\hat{N}_4\neq3n$ respectively. Solving the sum (\ref{c2}) we get
\begin{eqnarray}\label{c4}
G= \begin{cases} \hat{N}_4 & \hat{N}_4\neq3n   \\
4\hat{N}_4+6-3\delta_{\hat{N}_4,3} & \hat{N}_4=3n   \end{cases}.
\end{eqnarray}
The anti generations correspond to solutions satisfying,
\begin{eqnarray}\label{c5}
 k_1+k_2+k_3=0 \text{ mod } 3, \hspace{0.15in}  k_4-k_2=0 \text{ mod } h.
\end{eqnarray}
Where $h=1,3$ is the gcd of $\hat{N}_2$ and $\hat{N}_4$. As in the previous
case the anti generation number is given by
\begin{eqnarray}\label{c6}
\bar{G}=\sum^{2}_{i=1}\left(\left\lfloor\frac{G_{i}-1-a_i}{h}\right\rfloor+1\right).
\end{eqnarray}
Here, $G_i=G_{0,0,0},G_{1,1,1}$ and $a_i=k_2$ mod $h$. Finally the anti
generation number is given by
\begin{eqnarray}\label{c4}
\bar{G}= \begin{cases} \hat{N}_4 & \hat{N}_4\neq3n   \\
\frac{\hat{N}_4}{3} & \hat{N}_4=3n   \end{cases}.
\end{eqnarray}
Thus the net generation number is different from three for the $\{1,1,1,N_4\}$ theory.

\subsection{The $\{1,N_2,N_3\}$ Theory}

The last theory actually involves four cases with $N_2=1,2,3,4$. These are
solved using the technics presented in the previous cases, we state the
results for these cases. We first note that for $N_2=4$ the central charge of
the first two minimal models is three as in the last two theories. Indeed, the
solution for this case is similar, the only difference is that here solutions
satisfy $2k_1+k_2=0$ mod $M_2$. Thus $G_{k_1,k_2}$ is given by
\begin{eqnarray}\label{f1}
G_{k_1,k_2}=\delta_{2k_1+k_2,nM_2}\left( \hat{N}_3(1-\frac{2k_1+k_2}{6})\pm1\right).
\end{eqnarray}
Clearly here $M_2=1,2,3,6$ if we sum over $k_1$ and $k_2$ we find the
generation number is given by
\begin{eqnarray}\label{f2}
G= \begin{cases} 5\hat{N}_3+8-\delta_{\hat{N}_3,6} & \hat{N}_3=0 \text{ mod }6   \\
3\hat{N}_3+4 & \hat{N}_3=3 \text{ mod }6 \\
2\hat{N}_3+2 & \hat{N}_3=2,4 \text{ mod }6 \\
\hat{N}_3 & \hat{N}_3=1,5 \text{ mod }6  \end{cases}.
\end{eqnarray}
Solutions correspond to anti generation if and only if they satisfy
\begin{eqnarray}\label{f3}
 2k_1+k_2=0 \text{ mod } 6, \hspace{0.15in}  k_3-k_2=0 \text{ mod } h.
\end{eqnarray}
and again the anti generation number is given by the sum (\ref{b8}) with
$G_i=G_{0,0},G_{1,4}$, we find
\begin{eqnarray}\label{f4}
\bar{G}= \begin{cases} \frac{\hat{N}_3}{6} & \hat{N}_3=0 \text{ mod }6   \\
\frac{\hat{N}_3}{3} & \hat{N}_3=3 \text{ mod }6 \\
\frac{\hat{N}_3}{2}+1 & \hat{N}_3=2,4 \text{ mod }6 \\
\hat{N}_3 & \hat{N}_3=1,5 \text{ mod }6  \end{cases}.
\end{eqnarray}
Next, we turn to the first three cases. For these
theories we can write
\begin{eqnarray}\label{e1}
6\hat{N}_2=\alpha ndc, \hspace{0.2in}  \hat{N}_3=nde, \hspace{0.2in} \hat{N}_4=nce, \hspace{0.2in} g=\alpha nd^2,
\end{eqnarray}
where every pair of $c$,$e$ and $d$ is strange and $\alpha=1,2,3$ for
$\hat{N}_2=3,4,5$ respectively. If we define
\begin{eqnarray}\label{e2}
b^{\hat{N}_2}_{k_1,k_2}= \begin{cases} k_1(\hat{N}_2-1)+k_2(\hat{N}_1-1)  & \hat{N}_2=3    \\
k_1\hat{N}_2+k_2\hat{N}_1  & \hat{N}_2=4 \\
2k_1\hat{N}_2+2k_2\hat{N}_1  & \hat{N}_2=5  \end{cases},
\end{eqnarray}
solutions correspond to the condition
\begin{eqnarray}\label{e2}
k_3=b^{\hat{N}_2}_{k_1,k_2} \hspace{0.15in} \text{mod } d.
\end{eqnarray}
We can use $b^{\hat{N}_2}_{k_1,k_2}$ to write the lowest $k_3$ which satisfies this
condition as $a^{\hat{N}_2}_{k_1,k_2}=b^{\hat{N}_2}_{k_1,k_2}$ mod $d$. The generation
number is then given by,
\begin{eqnarray}\label{e4}
G=\sum^{N_1}_{k_1=0}\sum^{N_2}_{k_2=0}\left\lfloor\frac{\hat{N}_3}{d}(1-\frac{k_1}{\hat{N}_1}-
\frac{k_2}{\hat{N}_2}-\frac{2\delta_{k_1+k_2,0}}{\bar{N_3}})
-\frac{a^{\hat{N}_2}_{k_1,k_2}}{d}+1\right\rfloor-\left\lfloor\frac{6}{\hat{N}_3}\right\rfloor\delta_{\hat{N}_2,\hat{N}_3},
\end{eqnarray}
To find the anti generation number we define $h$ as the gcd of $\hat{N}_1$
and $\hat{N}_2$. Solutions which satisfy
\begin{eqnarray}\label{e5}
k_2-k_1=0 \hspace{0.08in} \text{mod }h,  \hspace{0.2in}
b^{\hat{N}_2}_{k_1,k_2}=0 \hspace{0.08in} \text{mod }2,  \hspace{0.2in}
k_3=b^{\hat{N}_2}_{k_1,k_2} \hspace{0.08in} \text{mod }nd,
\end{eqnarray}
can be shown to lie in the $\hat{Q}+K$ lattice. The anti generation number is then given
by
\begin{eqnarray}\label{e6}
\bar{G}=\sum^{N_1}_{k_1=0}\sum^{N_2}_{k_2=0}\delta_{b,2l}\delta_{k_2-k_1,sh}\left\lfloor\frac{\hat{N}_3}{nd}(1-\frac{k_1}{\hat{N}_1}-\frac{k_2}{\hat{N}_2}
-\frac{2\delta_{k_1+k_2,0}}{\bar{N_3}})
-\frac{\bar{a}^{\hat{N}_2}_{k_1,k_2}}{nd}+1\right\rfloor,
\end{eqnarray}
where we have suppressed $b$ indices and defined
$\bar{a}^{\hat{N}_2}_{k_1,k_2}=b^{\hat{N}_2}_{k_1,k_2}$ mod $nd$. The sums in $G$ and
$\bar{G}$ involve a small number of terms thus they provide an elegant solution. Finally,
we state that these results imply that the net number of generation is either zero or
bigger than three for the $\{1,N_2,N_3\}$ theory.

%%% ---------------------------------------------------------------------------

\def\baselinestretch{1}

\section{Discussion}

\def\baselinestretch{1.66}

\smallskip

In this work, we have described the construction of new $D=4$ heterotic string theories.
These string theories are achieved by means of the Gepner construction and the heterotic
string map \cite{db3}. In four dimensions, the Gepner construction requires a $c=9$, $N=2$
SCFT. New such SCFT were build by tensoring $r$ minimal and one nonstandard $N=2$ SCFT.
These SCFT are labeled by a set of positive integers $\{N_{r}\}$ and satisfy the central
charge equation,
\begin{eqnarray*}\label{dis1}
  9=\frac{3N}{2m-N}+\sum^r_{i=1}\frac{3N_i}{N_i+2},
\end{eqnarray*}
which determines $(m,N)$. All the solutions of this equation were found and are given by,
\begin{eqnarray*}\label{dis2}
\{N_r\}=\begin{cases} \{N_1\} & N_1 \in Z^+   \\
\{N_1,N_2\} & N_1,N_2 \in Z^+  \\
\{1,N_2,N_3\} & N_2\leq 4,\; N_3 \in Z^+  \\
\{2,2,N_3\} & N_3 \in Z^+ \\
\{1,1,1,N_4\} & N_4 \in Z^+  \end{cases},
\end{eqnarray*}
 and the table given in the appendix. It follows that all string theories with an internal SCFT comprised of one
nonstandard and $r$ minimal $N=2$ SCFT were found.\\
The heterotic theories in four dimensions have a gauge group which includes $E_8\times
E_6$. The massless spectrum includes some chiral fermions in the representation $27$ of
$E_6$ (generations) and some chiral fermions in the $\bar{27}$ of $E_6$
(anti–generations). The generation number corresponds to solutions of the $U(1)$ equation
(\ref{db5.9}), while the anti generation number corresponds to solutions of the $U(1)$
equation which lay on the $\hat{Q}+k$ lattice. By studying the $U(1)$ equation and
$\hat{Q}+k$ lattice the generation number and anti generation number for all the new
string theories were found and are given in section ($3$) and the appendix. Notably we found that only the
$\{6\}(13,18)$ theory predicts three net particle generation as seen in nature.\\
It was conjectured in ref. \cite{db3} that all the $N=2$ string theories correspond to
compactification on some Calabi–Yau manifold \cite{db6}. The Euler number of these
manifold is given by
\begin{eqnarray*}\label{dis1}
   \chi=G-\bar{G},
\end{eqnarray*}
and is thus found for all the string theories discussed. An interesting question is which
manifolds correspond to the new string theories. Of particular interest is the manifold
corresponding to the $\{6\}(13,18)$ theory and whether we can construct a realistic
string theory on this manifold. Such a theory is guaranteed to predict three net
generation. However, for a realistic description many other aspects of such a theory
should be studied. For example it should be possible to break the theory gauge group and
get the standard model gauge group $SU(3)\times SU(2)\times U(1)$. Finally, here we have
not dealt with string theories constructed by tensoring $r$ minimal and $2$ nonstandard
$N=2$ SCFT. It is evident from the discussion in section ($3.3$) that an infinite number of
such $D=4$ consistent string theories exist. The net generation number and the questions
we have posed are equally interesting for these theories as well.

%--------------------------------------------------------------------------%
\newpage
\renewcommand\thesection{}
\section{Appendix}

\hspace{0.2in} Generation and anti generation numbers for the models studied numerically.
\begin{equation*}\scriptsize
\begin{array}{||c|c|c|c|c|} \hline \hline
 \{N_r\} & \frac{m}{N}  & G & \bar{G} & G_{net} \\     \hline  \hline
 \{0\} & \frac{2}{3} & 1 & 0 & 0\\ \hline
 \{1,5,5\} & \frac{47}{52} & 45 & 7 & 38 \\                   \hline
 \{1,5,6\} & \frac{185}{202} & 50 & 28 & 22 \\                      \hline
 \{1,5,7\} & \frac{137}{148} & 56 & 20 & 36 \\                            \hline
 \{1,5,8\} & \frac{113}{121} & 34 & 34 & 0 \\                                   \hline
 \{1,5,9\} & \frac{493}{524} & 68 & 68 & 0 \\                                         \hline
 \{1,5,10\} & \frac{89}{94} & 74 & 13 & 61 \\                                               \hline
 \{1,5,11\} & \frac{575}{604} & 79 & 79 & 0 \\                        \hline
 \{1,5,12\} & \frac{22}{23} & 47 & 6 & 41 \\                         \hline
 \{1,5,13\} & \frac{73}{76} & 33 & 33 & 0 \\                          \hline
 \{1,5,14\} & \frac{349}{362} & 96 & 51 & 45 \\                       \hline
 \{1,5,15\} & \frac{739}{764} & 102 & 102 & 0 \\                     \hline
 \{1,5,16\} & \frac{65}{67} & 57 & 19 & 38 \\       \hline
 \{1,5,17\} & \frac{821}{844} & 113 & 113 & 0 \\          \hline
 \{1,5,18\} & \frac{431}{442} & 119 & 62 & 57 \\                \hline
 \{1,5,19\} & \frac{43}{44} & 130 & 6 & 124 \\                        \hline
 \{1,5,20\} & \frac{236}{241} & 69 & 69 & 0 \\                              \hline
 \{1,5,21\} & \frac{985}{1004} & 137 & 137 & 0 \\                                 \hline
 \{1,5,22\} & \frac{57}{58} & 49 & 25 & 24 \\                                           \hline
 \{1,5,23\} & \frac{1067}{1084} & 148 & 148 & 0 \\                                            \hline
 \{1,5,24\} & \frac{277}{281} & 80 & 80 & 0 \\               \hline
 \{1,5,25\} & \frac{383}{388} & 161 & 55 & 106 \\            \hline
 \{1,5,26\} & \frac{85}{86} & 167 & 12 & 155 \\              \hline
 \{1,5,27\} & \frac{1231}{1244} & 171 & 171 & 0 \\           \hline
 \{1,5,28\} & \frac{106}{107} & 91 & 30 & 61 \\              \hline
 \{1,5,29\} & \frac{1313}{1324} & 182 & 182 & 0 \\           \hline
 \{1,5,30\} & \frac{677}{682} & 189 & 97 & 92 \\             \hline
 \{1,5,31\} & \frac{155}{156} & 67 & 67 & 0 \\                     \hline
 \{1,5,32\} & \frac{359}{361} & 103 & 103 & 0 \\                   \hline
 \{1,5,33\} & \frac{211}{212} & 209 & 30 & 179 \\                  \hline
 \{1,5,34\} & \frac{253}{254} & 211 & 36 & 175 \\    \hline
 \{1,5,35\} & \frac{1559}{1564} & 217 & 217 & 0 \\         \hline
 \{1,5,36\} & \frac{400}{401} & 114 & 114 & 0 \\                 \hline
 \{1,5,37\} & \frac{547}{548} & 229 & 78 & 151 \\                      \hline
 \{1,5,38\} & \frac{841}{842} & 234 & 120 & 114 \\                           \hline
 \{1,5,39\} & \frac{1723}{1724} & 240 & 240 & 0  \\                                \hline
 \{1,6,6\} & \frac{13}{14} & 30 & 5 & 25 \\                                              \hline
 \{1,6,7\} & \frac{77}{82} & 63 & 13 & 50 \\                    \hline
 \{1,6,8\} & \frac{127}{134} & 70 & 22 & 48 \\                  \hline \hline
\end{array}
\begin{array}{||c|c|c|c|c||} \hline \hline
 \{N_r\} & \frac{m}{N}  & G & \bar{G} & G_{net}\\  \hline \hline
 \{1,6,9\} & \frac{277}{290} & 76 & 43 & 33 \\                                 \hline
 \{1,6,10\} & \frac{25}{26} & 86 & 4 & 82 \\                     \hline
 \{1,6,11\} & \frac{323}{334} & 90 & 51 & 39 \\                  \hline
 \{1,6,12\} & \frac{173}{178} & 97 & 29 & 68 \\                  \hline
 \{1,6,13\} & \frac{41}{42} & 37 & 21 & 16 \\  \hline
 \{1,6,14\} & \frac{49}{50} & 111 & 8 & 103 \\   \hline
 \{1,6,15\} & \frac{415}{422} & 115 & 66 & 49 \\       \hline
 \{1,6,16\} & \frac{73}{74} & 123 & 12 & 111 \\                   \hline
 \{1,6,17\} & \frac{461}{466} & 128 & 73 & 55 \\                             \hline
 \{1,6,18\} & \frac{121}{122} & 136 & 20 & 116 \\                                       \hline
 \{1,6,19\} & \frac{169}{170} & 142 & 28 & 114 \\          \hline
 \{1,6,20\} & \frac{265}{266} & 148 & 44 & 104 \\          \hline
 \{1,6,21\} & \frac{553}{554} & 154 & 88 & 66  \\              \hline
 \{1,7,7\} & \frac{19}{20} & 77 & 3 & 74 \\                  \hline
 \{1,7,8\} & \frac{47}{49} & 44 & 16 & 28 \\                       \hline
 \{1,7,9\} & \frac{205}{212} & 86 & 32 & 54 \\                           \hline
 \{1,7,10\} & \frac{37}{38} & 95 & 6 & 89 \\                                   \hline
 \{1,7,11\} & \frac{239}{244} & 100 & 37 & 63 \\                                     \hline
 \{1,7,12\} & \frac{64}{65} & 58 & 21 & 37 \\                                              \hline
 \{1,7,13\} & \frac{91}{92} & 117 & 15 & 102 \\            \hline
 \{1,7,14\} & \frac{145}{146} & 122 & 24 & 98 \\                 \hline
 \{1,7,15\} & \frac{307}{308} & 129 & 48 & 81 \\                       \hline
 \{1,8,8\} & \frac{31}{32} & 88 & 5 & 83 \\                                  \hline
 \{1,8,9\} & \frac{169}{173} & 52 & 52 & 0 \\                                      \hline
 \{1,8,10\} & \frac{61}{62} & 103 & 10 & 93 \\                                           \hline
 \{1,8,11\} & \frac{197}{199} & 61 & 61 & 0 \\                        \hline
 \{1,8,12\} & \frac{211}{212} & 118 & 35 & 83 \\                        \hline
 \{1,9,9\} & \frac{67}{68} & 108 & 10 & 98 \\                           \hline
 \{1,9,10\} & \frac{133}{134} & 111 & 20 & 91 \\            \hline
 \{1,9,11\} & \frac{859}{860} & 120 & 120 & 0  \\             \hline
 \{2,3,3\} & \frac{23}{26} & 31 & 5 & 26 \\                  \hline
 \{2,3,4\} & \frac{67}{74} & 37 & 14 & 23 \\                  \hline
 \{2,3,5\} & \frac{153}{166} & 43 & 28 & 15 \\                \hline
 \{2,3,6\} & \frac{43}{46} & 49 & 9 & 40 \\                    \hline
 \{2,3,7\} & \frac{191}{202} & 53 & 35 & 18 \\                \hline
 \{2,3,8\} & \frac{21}{22} & 62 & 4 & 58 \\                            \hline
 \{2,3,9\} & \frac{229}{238} & 65 & 43 & 22 \\               \hline
 \{2,3,10\} & \frac{31}{32} & 36 & 13 & 23 \\               \hline
 \{2,3,11\} & \frac{267}{274} & 75 & 50 & 25 \\   \hline           \hline
\end{array}
\begin{array}{|c|c|c|c|c||} \hline  \hline
 \{N_r\} & \frac{m}{N}  & G & \bar{G} & G_{net} \\    \hline      \hline
 \{2,3,12\} & \frac{143}{146} & 82 & 29 & 53 \\              \hline
 \{2,3,13\} & \frac{61}{62} & 89 & 12 & 77 \\                              \hline
 \{2,3,14\} & \frac{81}{82} & 92 & 16 & 76 \\                        \hline
 \{2,3,15\} & \frac{343}{346} & 97 & 65 & 32 \\                \hline
 \{2,3,16\} & \frac{181}{182} & 103 & 36 & 67 \\         \hline
 \{2,3,17\} & \frac{381}{382} & 108 & 72 & 36  \\  \hline
 \{2,4,4\} & \frac{13}{14} & 47 & 3 & 44 \\                       \hline
 \{2,4,5\} & \frac{89}{94} & 51 & 20 & 31 \\                      \hline
 \{2,4,6\} & \frac{25}{26} & 59 & 6 & 53 \\                        \hline
 \{2,4,7\} & \frac{37}{38} & 65 & 8 & 57 \\                        \hline
 \{2,4,8\} & \frac{61}{62} & 71 & 15 & 56 \\                       \hline
 \{2,4,9\} & \frac{133}{134} & 76 & 30 & 46 \\                     \hline
 \{2,5,5\} & \frac{29}{30} & 61 & 6 & 55 \\                          \hline
 \{2,5,6\} & \frac{57}{58} & 65 & 12 & 53 \\                      \hline
 \{2,5,7\} & \frac{253}{254} & 72 & 48 & 24 \\                    \hline
 \{3,3,3\} & \frac{11}{12} & 44 & 2 & 42 \\                              \hline
\{3,3,4\} & \frac{16}{17} & 28 & 6 & 22 \\                                \hline
 \{3,3,5\} & \frac{73}{76} & 53 & 13 & 40 \\                            \hline
 \{3,3,6\} & \frac{41}{42} & 60 & 8 & 52 \\                                \hline
 \{3,3,7\} & \frac{91}{92} & 66 & 16 & 50 \\    \hline
 \{3,4,4\} & \frac{31}{32} & 53 & 6 & 47 \\           \hline
 \{3,4,5\} & \frac{106}{107} & 36 & 36 & 0  \\              \hline
 \{1,1,2,2\} & \frac{7}{8} & 14 & 4 & 10 \\                       \hline
 \{1,1,2,3\} & \frac{67}{74} & 30 & 14 & 16 \\                          \hline
 \{1,1,2,4\} & \frac{13}{14} & 38 & 3 & 35 \\                                 \hline
 \{1,1,2,5\} & \frac{89}{94} & 40 & 20 & 20 \\                                      \hline
 \{1,1,2,6\} & \frac{25}{26} & 47 & 6 & 41 \\                                             \hline
 \{1,1,2,7\} & \frac{37}{38} & 51 & 9 & 42 \\                                                   \hline
 \{1,1,2,8\} & \frac{61}{62} & 56 & 14 & 42 \\                                                        \hline
 \{1,1,2,9\} & \frac{133}{134} & 60 & 30 & 30 \\                                                            \hline
 \{1,1,3,3\} & \frac{16}{17} & 38 & 4 & 34 \\                                                                     \hline
 \{1,1,3,4\} & \frac{31}{32} & 43 & 8 & 35 \\                                                                           \hline
 \{1,1,3,5\} & \frac{106}{107} & 48 & 24 & 24 \\                                                                              \hline
 \{1,2,2,2\} & \frac{13}{14} & 33 & 3 & 30 \\                                                                                       \hline
 \{1,2,2,3\} & \frac{31}{32} & 21 & 13 & 8 \\                                                                                             \hline
 \{1,1,1,1,1\} & \frac{7}{8} & 26 & 1 & 25 \\                                                                                                   \hline
 \{1,1,1,1,2\} & \frac{13}{14} & 31 & 2 & 29 \\                                                                                                       \hline
 \{1,1,1,1,3\} & \frac{31}{32} & 36 & 4 & 32 \\                                                                                                             \hline
 & & & & \\ \hline      \hline
\end{array} \hspace{1in}
\end{equation*}

\def\baselinestretch{1.66}

\smallskip

\bibliographystyle{amsplain}

%-----------------------------------------------------------------------------

\end{document}